\documentclass[useAMS,usenatbib,usegraphicx]{mn2e}
\usepackage[bw]{optional}

\title[The Spine of the X-ray L-T relation] {Nature versus Nurture:
 The curved spine of the galaxy cluster X-ray luminosity -- temperature relation}
\author[W. G. Hartley et al.]  {W.~G.~Hartley,$^1$ L.~Gazzola,$^1$ F.~R.~Pearce,$^1$ S.~T.~Kay$^{2,3}$ \& P.~A.~Thomas$^4$\\
$^{1}$School of Physics and Astronomy, University of Nottingham,
Nottingham, NG7~2RD, UK\\ 
$^2$Jodrell Bank Centre for Astrophysics, School of Physics and Astronomy, The University of Manchester, Manchester M13 9PL\\ 
$^3$Astrophysics, Department of Physics, University of Oxford, Keble Road, Oxford OX1 3RH\\
$^4$Department of Physics and Astronomy, University of Sussex, Falmer, Brighton BN1 9QH\\} 

\date{Released 2007}

\pagerange{\pageref{firstpage}--\pageref{lastpage}} \pubyear{2007}

\begin{document}

\label{firstpage}

\maketitle

\begin{abstract}

The physical processes that define the spine of the galaxy cluster
X-ray luminosity -- temperature (L-T) relation are investigated using
a large hydrodynamical simulation of the Universe. This simulation
models the same volume and phases as the Millennium Simulation and has
a linear extent of $500h^{-1}$Mpc. We demonstrate that mergers
typically boost a cluster along but also slightly below the L-T
relation. Due to this boost we expect that all of the very brightest
clusters will be near the peak of a merger. Objects from near the top
of the L-T relation tend to have assembled much of their mass earlier
than an average halo of similar final mass. Conversely, objects from
the bottom of the relation are often experiencing an ongoing or recent
merger. 

\end{abstract}


\section{Introduction}

Since the launch of the XMM-Newton and Chandra satellites
\citep{XMM,Chandra}, measurements of the X-ray emission from hot gas
in clusters of galaxies have achieved unprecedented levels of accuracy
and depth. However, the physical origin of the scaling relations between
observable quantities, such as the luminosity of the X-ray emitting
gas and its temperature, remain only partly understood.

There are currently a number of surveys \citep{Romer, Schwope,
Pierre} in progress with the potential to greatly expand our
understanding of the processes that define correlations such as the
luminosity-temperature (L-T) relation of clusters. For this potential
to be realised we require a sound theoretical basis upon which to
work. To this end, numerical hydrodynamical simulations have become
indispensable tools and continue to grow in size and complexity
\citep{Pearce, Kay07, Faltenbacher} but they have to date lacked
a sufficiently large dynamic range in mass. In this work we use a
hydrodynamical model of a large volume that contains over a hundred galaxy
clusters. For the first time we are able to study the evolutionary
processes within a cosmological context as we have hundreds of well
resolved objects spanning a large dynamic range 
rather than the more typical handful \citep{Rowley} (hereafter R04), 
or idealised models \citep{Ritchie,Poole}.

This paper is organised as follows: in the remainder of this section
we summarise the work done to date on defining the physical processes
that define the shape of the L-T relation. Then, in section
2, we give an account of the simulations we have undertaken, explain
how our cluster sub-sample was selected and how the properties of
these clusters were derived.  Section 3 details our results before we
discuss their implications and conclude in section 4.

X-rays are chiefly emitted from the hot gas in clusters via thermal
bremsstrahlung (for dark matter halos more massive than $10^{14}
h^{-1} {\rm M_\odot}$ their temperature is typically above $2{\rm
keV}$). For such a homologous population \citet{Kaiser} showed that
simple scaling relations were expected between bulk properties such as
the mass, temperature and luminosity.  Observational work subsequently
found that the properties of X-ray clusters where indeed related but
the slopes of the relations were not those derived by Kaiser.

\begin{figure*}
\begin{minipage}{150mm}
\begin{center}
\opt{bw}{
\includegraphics[angle=0, width=\textwidth]{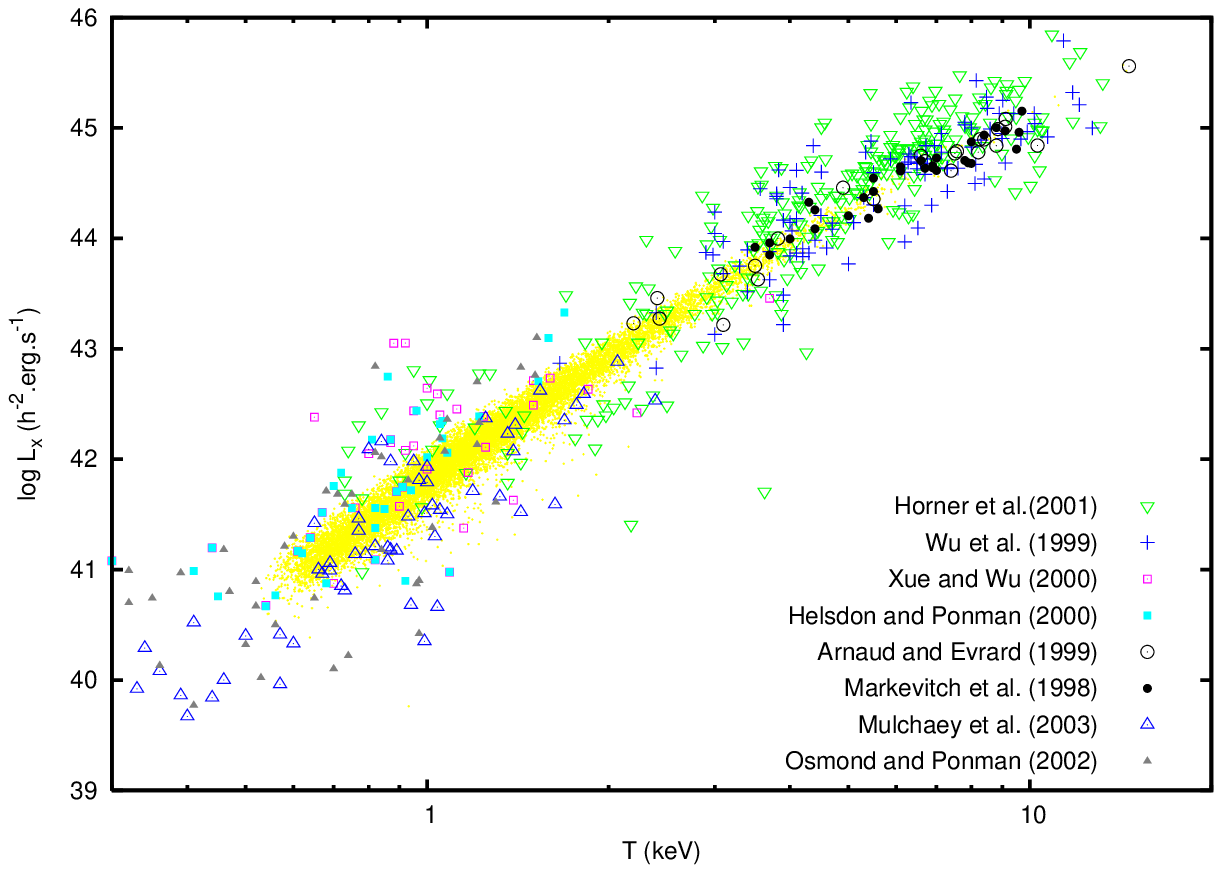}
}
\opt{col}{
\includegraphics[angle=0, width=\textwidth]{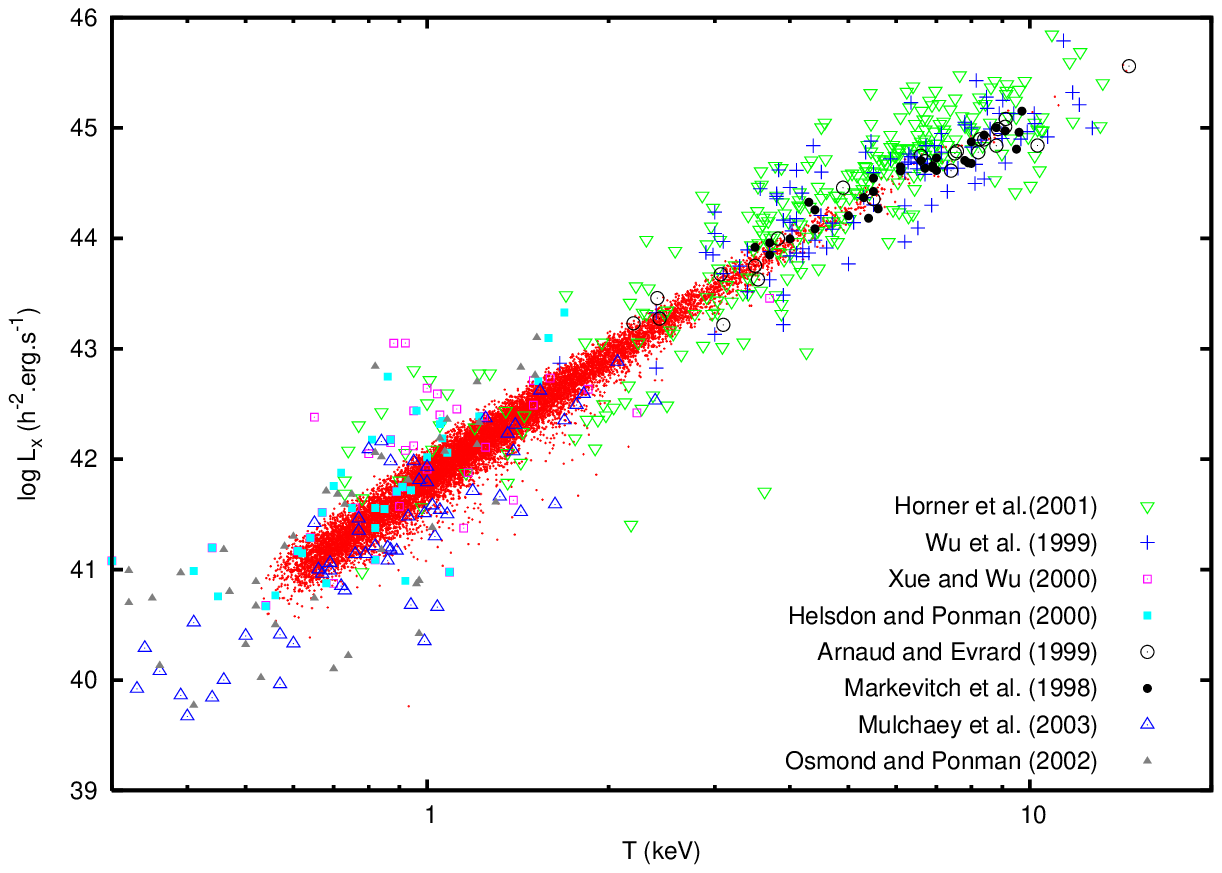}
}
\caption{
Compilation of low-redshift observed group and cluster X-ray
luminosities within $r_{500}$ compared to their emission-weighted
temperature. $r_{500}$ is a radius enclosing an overdensity of
500. The small points are the simulated groups and clusters used in
this work. The data was taken from variously: \citet{Markevitch,Arnaud,Wu99,
Helsdon,Xue,Horner,Mulchaey,Osmond}.}
\label{obsdata}
\end{center}
\end{minipage}
\end{figure*}

\citet{Kaiser} assumed that galaxy clusters were self-similar entities
and that therefore only a single property, such as the mass, was
required in order to describe the other bulk properties. Such a
homology results in an L-T relation with a power-law slope of
2. However, as figure~\ref{obsdata} demonstrates, X-ray observations
of clusters with a median redshift of $\sim 0.07$
found that the slope was closer to 3 \citep{Markevitch,Arnaud,Wu99,
Xue,Horner,Mulchaey, Osmond} and perhaps became even steeper
on group scales \citep{Helsdon}.

Hydrodynamical simulations performed in the absence of cooling or any
additional heat sources other than compression and shock heating have
long been known to reproduce the self-similar hierarchy well
\citep{ENF,NFW}. Unfortunately they do not reproduce either the slope
or the normalisation of the observations, producing clusters that are
too bright for any given temperature, even at the bright end.
Following this work, simulations with limited physics within a
cosmological volume have been used in an attempt to reconcile the
apparent discrepancy between theory and observation regarding the
slope of the L-T relation \citep{Pearce, Muanwong,
Bialek,Borgani02}. These models showed that a simple cooling or
preheating scheme was sufficient to match the simulated L-T relation
to that observed at redshift zero.  More recently \citet{Kay07} investigated 
the effects of feedback on the X-ray properties of clusters in hydrodynamical 
simulations, and demonstrated that their results were in good agreement with 
both the observed scaling relations and structural properties (e.g. entropy 
and temperature profiles), particularly for cool-core clusters.

\citet{Balogh06} investigated the role that preheating, cooling and
concentration of the halo profile can have on the scaling
relations. They found that, for a realistic range of halo
concentrations, the scatter generated was minimal in comparison with
observed values. Variations in the cooling time of the gas in the
centre of clusters could account for much of the scatter but is
limited by the age of the universe and so could not explain the whole
range. Finally, varying feedback from supernovae and AGN could explain
the entire range, but required an order of magnitude difference in
energy injection to cover the whole envelope.  Their result implies
that it is processes in the cores of clusters that are primarily
responsible for driving the scatter in the scaling relations. This
confirms earlier work by \citet{Fabian, Markevitch} and
\citet{McCarthy}.  \citet{Kay07} identify the scatter with strong cool
core clusters, and expect the scatter to be smaller at high redshift
due to the diminished prevalence of such systems.  Nowadays, the
general consensus is that the scatter is largely due to the strength
of the X-ray core. In this work, which includes strong preheating,
X-ray cores are absent. This allows us to study the shape of the
relation without the additional complication of a large intrinsic
scatter.

In this work we will use a sample of halos identified from the full 
simulation volume. With these we will show that because mergers tend to move clusters
up the L-T relation they extend it beyond the point where the most
massive, relaxed clusters are expected to lie. Thus many of the
brightest, most luminous objects are ongoing or recent merger events
which (as R04 point out) may be difficult to resolve
observationally if they are close to the peak of the merger. In
addition, because we have many closely spaced outputs we can track the
motion of each of our clusters on the L-T plane, allowing us to define
a ``mean merger'' vector. As this vector is not perfectly
parallel to the L-T relation but rather falls slightly below it, a
gentle roll in the relation naturally arises. 

\section{The simulations}
\label{sims}

The simulation used in this work is part of the Millennium Gas
Simulations \citep{Pearce07}. This sequence of hydrodynamical
simulations all have the same volume as the Millennium Simulation
\citep{Springel} as well as utilising the same amplitude and phase for
the initial perturbations.  The cosmological parameters for both the
Millennium Simulation and the gas counterparts were: $\Omega_\Lambda =
0.75, \Omega_M = 0.25, \Omega_b = 0.045, h = 0.73, n = 1$, and
$\sigma_8 = 0.9$, where the Hubble constant is characterised as $100h
{\rm km s^{-1} Mpc^{-1}}.$ These cosmological parameters are
consistent with recent combined analyses from {\it WMAP} data
\citep{Spergel} and the 2dF galaxy redshift survey \citep{Colless}.
The simulation volume is a comoving cube of linear size $500
h^{-1}{\rm Mpc}$ containing $500$ million dark matter particles and
$500$ million gas particles. Their masses are $1.422\times
10^{10}h^{-1}{\rm M_\odot}$ and $3.12\times 10^{9}h^{-1}{\rm M_\odot}$
respectively. The simulation includes radiative cooling of the gas,
with the metalicity set at a constant value of $0.3 {\rm Z_{\odot}}$,
similar to that observed within the intra-cluster medium
\citep{Sarazin} and preheating. The preheating is implemented in a
similar way to \citet{Borgani}: at redshift $4$ the whole volume is
heated to $200 {\rm keV/cm^{2}}$, a value chosen such that the
resulting L-T relation at redshift $0$ matches observations. Star
particles are formed from cold, dense gas particles when a temperature
threshold ($2\times10^{4}$ K), a density threshold ($\rm{n}_H = 4.18 
\times 10^{-27}$ g.cm$^{-3}$ and an overdensity threshold ($100$ times 
critical) are all passed, but the process of converting gas to stars has 
no effect on the thermal dynamics of the system, other than to make the 
particles collisionless.  The effect of the preheating in this simulation is so
extreme as to prevent any further star formation since redshift $4$.

\subsection{Sample selection}
\label{sampleselect}

At redshift zero the entire volume was processed to obtain a set of
friends-of-friends halos with a linking length of $ 0.2$ times the mean
interparticle separation. Within each of these halos the most bound
particle was found and used as the centre for a spherical over-density
calculation that extended to $r_{200}$, a radius enclosing an
over-density of 200 times the cosmic mean. The analysis presented in
this work is for a fixed radius of $r_{500}$ (the radius enclosing an 
over-density 500 times the the cosmic mean density), roughly $0.59 \times
r_{200}$ for NFW halos \citep{NFW97} of typical concentration. Within
this radius we calculate the bolometric luminosity and the emission
weighted temperature assuming a standard \citep{Sutherland} cooling
function for a uniform metalicity gas of $0.3 {\rm Z_\odot}$.

Three sub-samples were selected from the top, bottom and median of the
L-T relation. We refer to the sample of clusters that are more
luminous and/or cooler than expected as coming from the top of the
relation, with conversely under luminous, hot clusters coming from the
bottom. We also select a control sample of clusters from close to the
median of the relation. The clusters were selected such that the range
of masses within each sample spanned the entirety of the available
relation. We only consider objects containing more than 1000 particles
and that are at least two virial radii away from any larger
neighbour. This ensures a meaningful estimate of the cluster bulk
properties.

\begin{figure}
\begin{center}
\opt{bw}{
\includegraphics[angle=0, width=250pt]{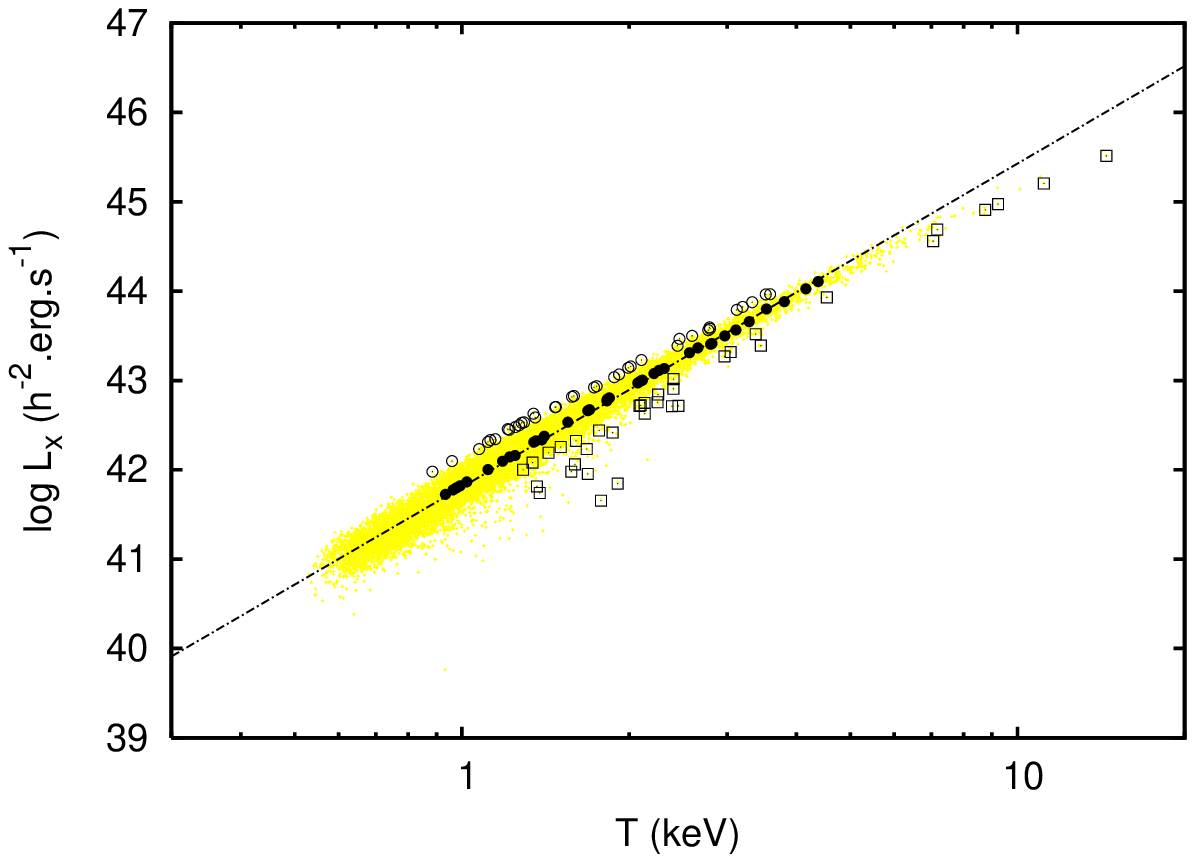}
}
\opt{col}{
\includegraphics[angle=0, width=250pt]{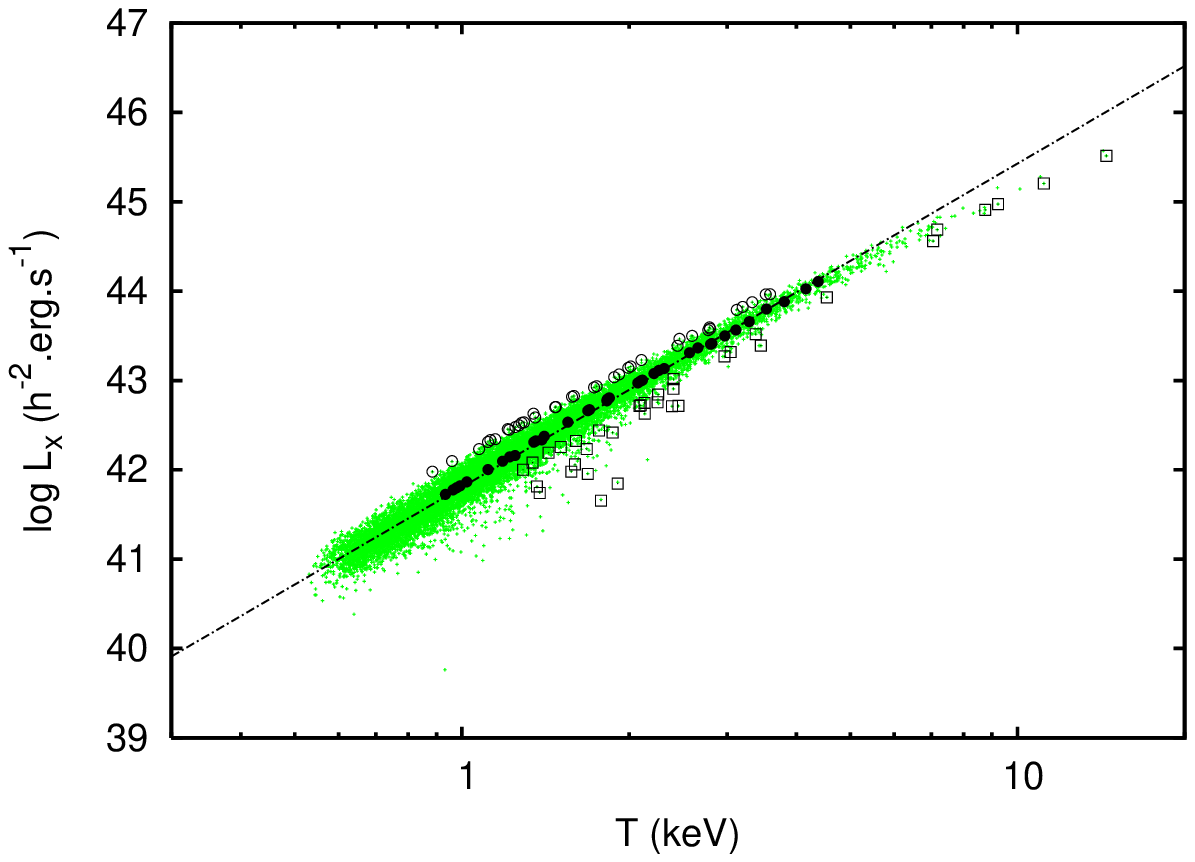}
}
\caption{The clusters selected for further analysis: the high and low scatter
 samples are shown as open circles and squares respectively, above and below the line which
 indicates our fitted mean relation. The control clusters are shown as
 solid circles. The underlying faint points show the full sample
 at redshift $0$. }
\label{sample}
\end{center}
\end{figure}

Once selected at redshift zero, each of our 108 clusters was traced
backwards in time until their mass dropped below our imposed
resolution threshold of 1000 particles.  The final locations of the
selected clusters on the L-T plane are shown on figure~\ref{sample},
where the high scatter clusters are indicated by open circles, low scatter by open squares  
and the control sample by filled circles. The 
full sample is shown faintly in the background, together with our
fitted median relation indicated by the line.  Once the mass
accretion histories of the sample had been extracted a small amount of
smoothing was introduced in order to remove merger induced ringing in
the cluster mass.

\begin{figure*}
\begin{center}
\opt{bw}{
\includegraphics[angle=0, width=15cm]{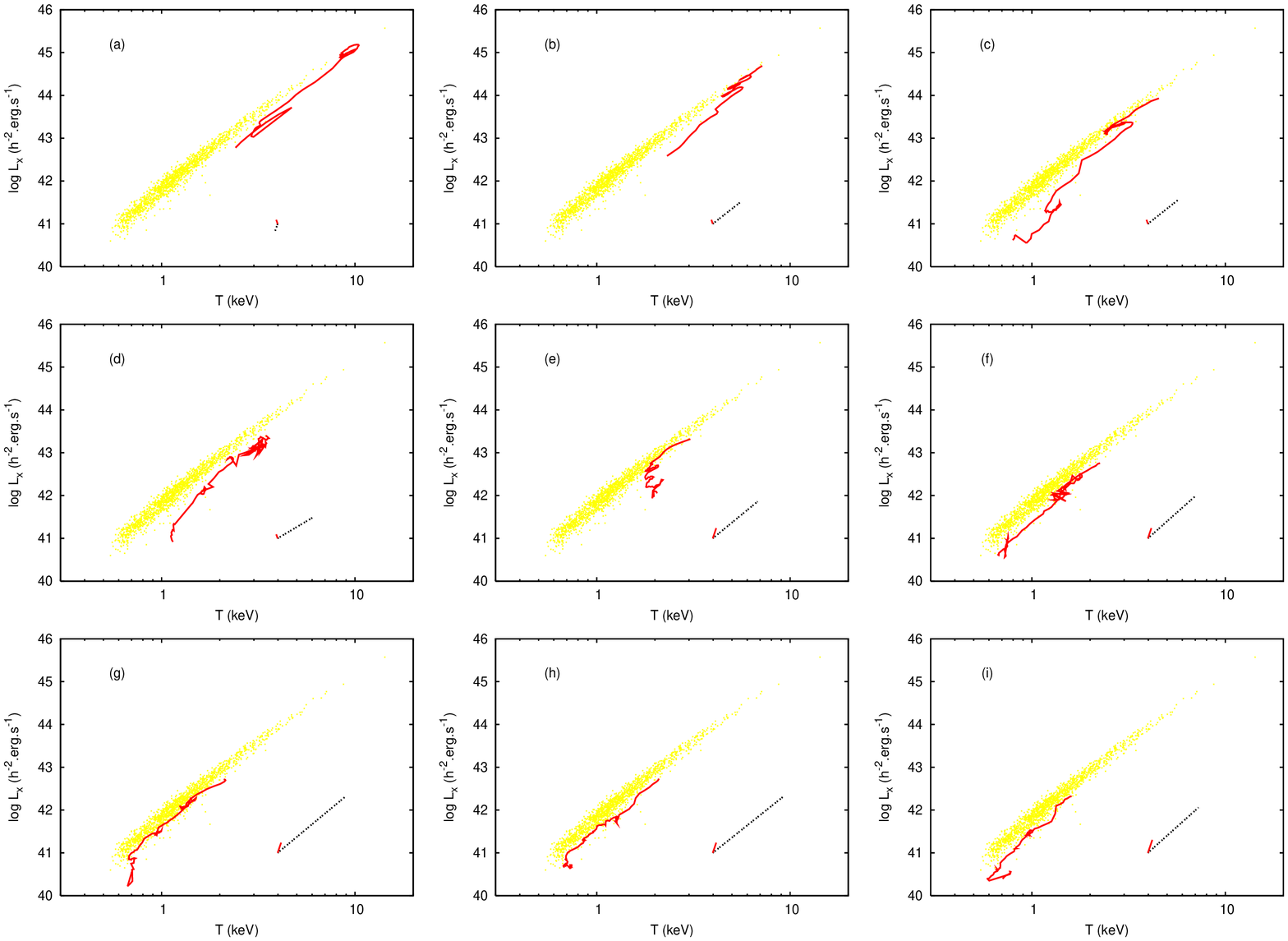}
}
\opt{col}{
\includegraphics[angle=0, width=15cm]{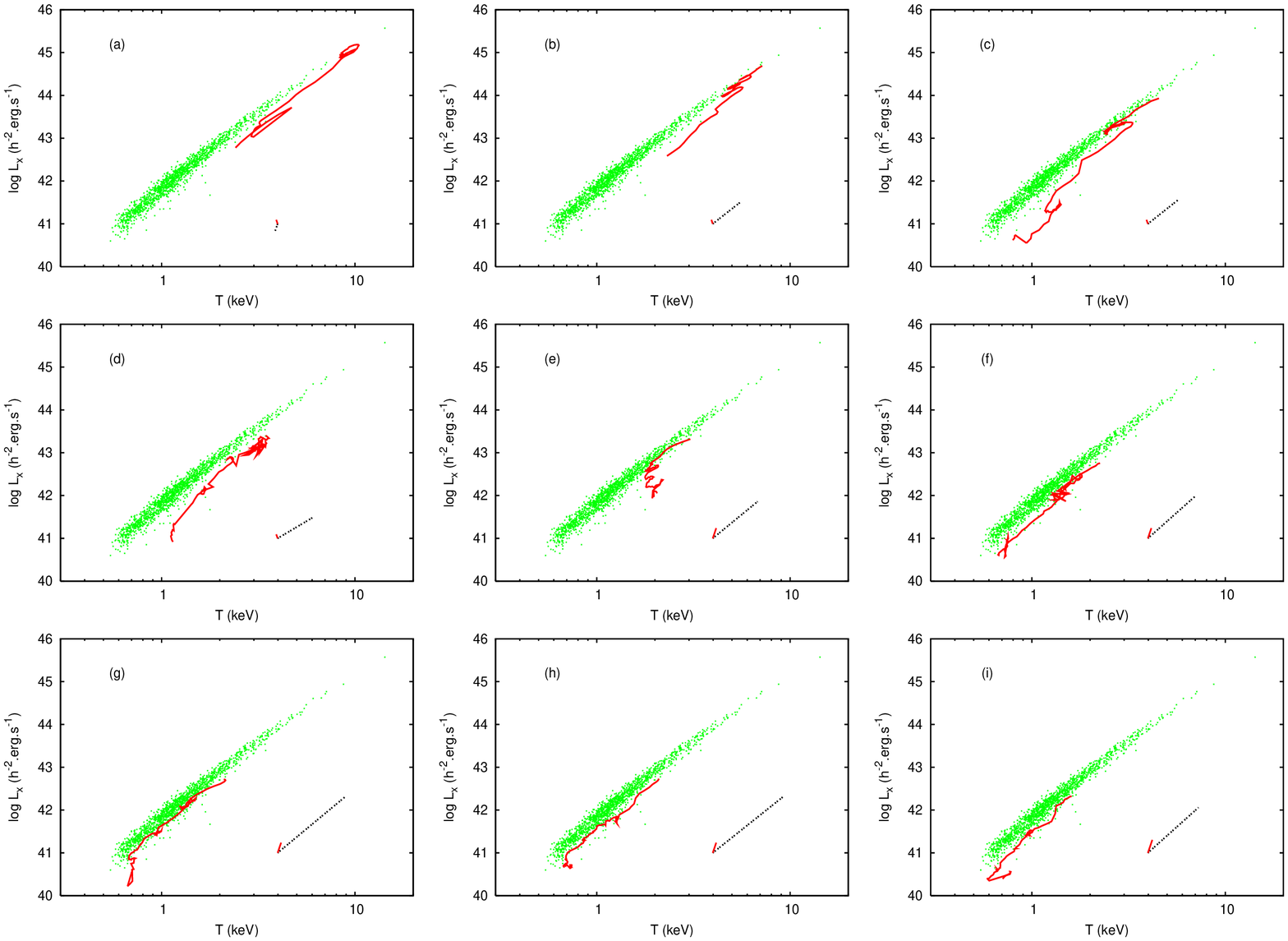}
}
\caption{X-ray evolution in the L-T plane of $9$ of the
low-scattered clusters ranked by final mass. In each of the nine frames the L-T relation at
redshift zero is defined by a sample of clusters (background points) whilst
the track showing the evolution of each cluster is shown as the dark
line. 
To the bottom right of each panel the evolution in the L-T plane for each
cluster (long dashed line) and the mean of the control sample
members of similar mass (short solid line)
between a redshift of $0.5$ and $0$.  }
\label{llt}
\end{center}
\end{figure*}

\begin{figure*}
\begin{center}
\opt{bw}{
\includegraphics[angle=0, width=15cm]{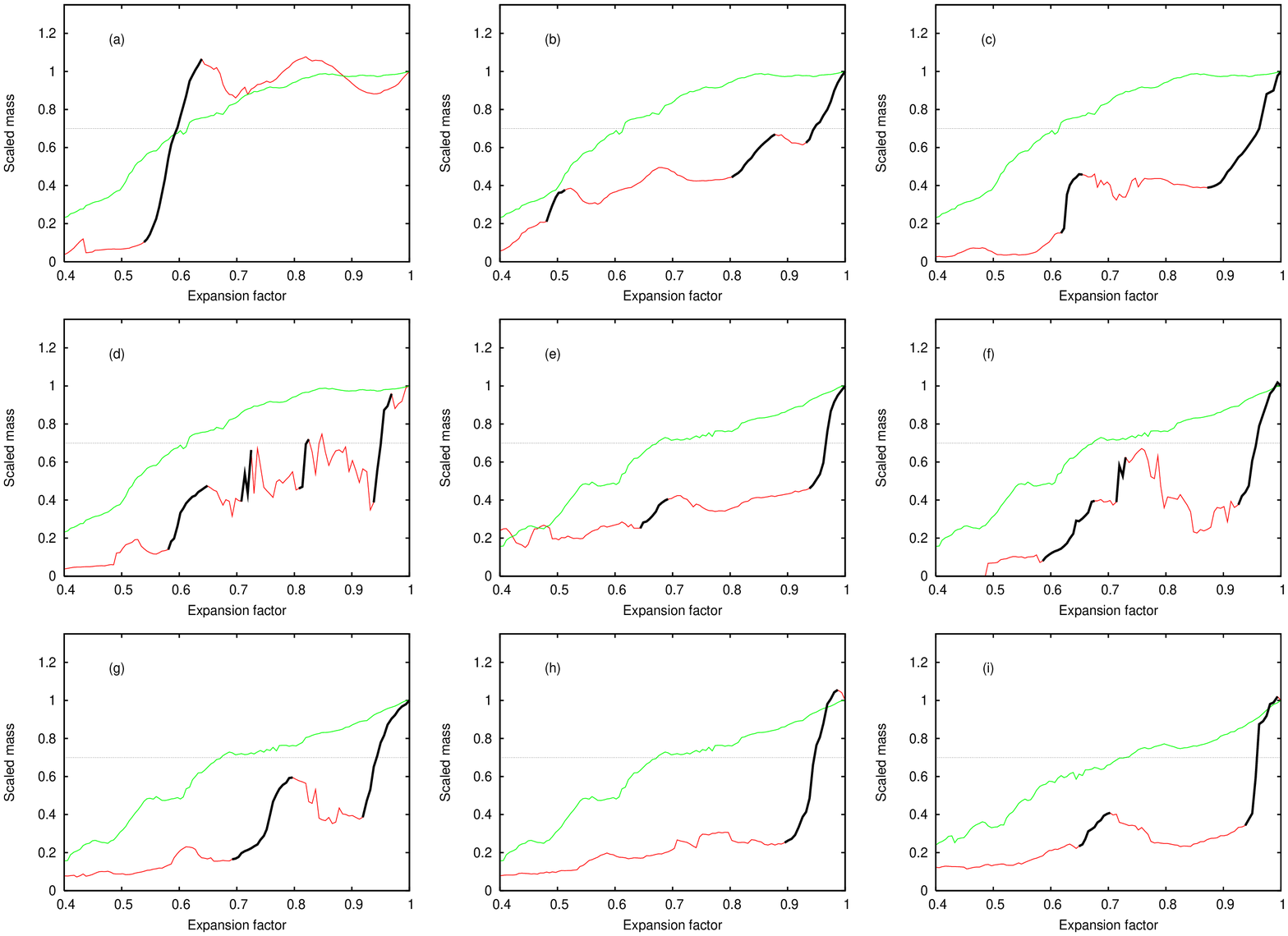}
}
\opt{col}{
\includegraphics[angle=0, width=15cm]{figs/masslow}
}
\caption{Mass accretion history for the 9 
low-scattered clusters whose L-T evolution was shown in figure~\ref{llt}. 
The lighter line is the mass accretion history
for the mean of similar mass clusters in the control sample 
whereas the darker line shows the history of the
individual cluster. Bold sections demote periods when the cluster is
undergoing a merger as defined in section 3.3 below. 
The mass of the cluster at each expansion factor
is normalised by its final mass, labelled as ``scaled mass''. Also
plotted is a horizontal dotted line to show when a cluster has
assembled 70\% of its final mass.  }
\label{lm}
\end{center}
\end{figure*}

\section{Halo properties}

\subsection{Low-scattered halos}

The L-T histories of nine low scattered halos ranked by final mass are
shown in figure~\ref{llt}, with their respective mass accretion
histories given in figure~\ref{lm}. The dark track in the L-T plane
follows the history of each object from redshift 1.5 to the present
day. The background points
show the location of the entire sample of clusters in the L-T plane at
$z=0$. To the bottom right of each panel are two additional
vectors. The short line shows the mean evolution of the control sample
between $z=0.5$ and the present. The other longer line shows the
evolution of each particular group over the same redshift
interval. While the control sample moves slightly up in luminosity and
down in temperature, 8 of the 9 low scattered clusters move
dramatically to larger luminosities and temperatures, nearly parallel
to the spine of the L-T relation. This is in agreement with the trend
noticed by R04.  Figure~\ref{lm}, which shows the corresponding mass
accretion histories for these objects demonstrates that 8 of the 9 low
scattered objects are in the process of an ongoing major merger and
are much hotter and brighter than would be typical for objects of
their mass. In each of these panels merger events are denoted by the
bold sections of line.
The masses at redshift 1.5 are significantly lower than
expected in all bar one case, (the mean mass accretion history for
objects of each mass is indicated by the dotted line on
figure~\ref{lm}).

\subsection{High-scattered halos}

\begin{figure*}
\begin{center}
\opt{bw}{
\includegraphics[angle=0, width=15cm]{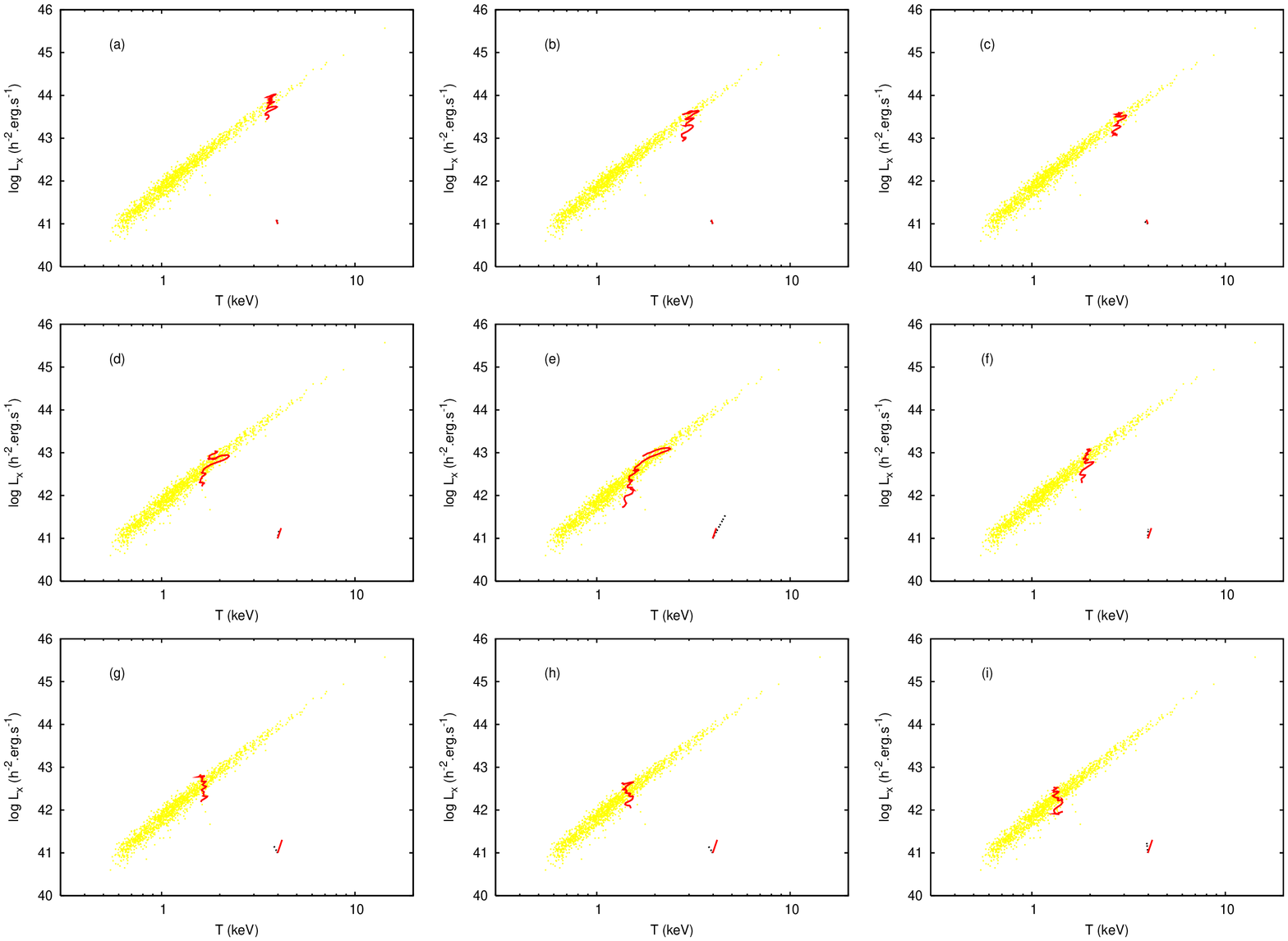}
}
\opt{col}{
\includegraphics[angle=0, width=15cm]{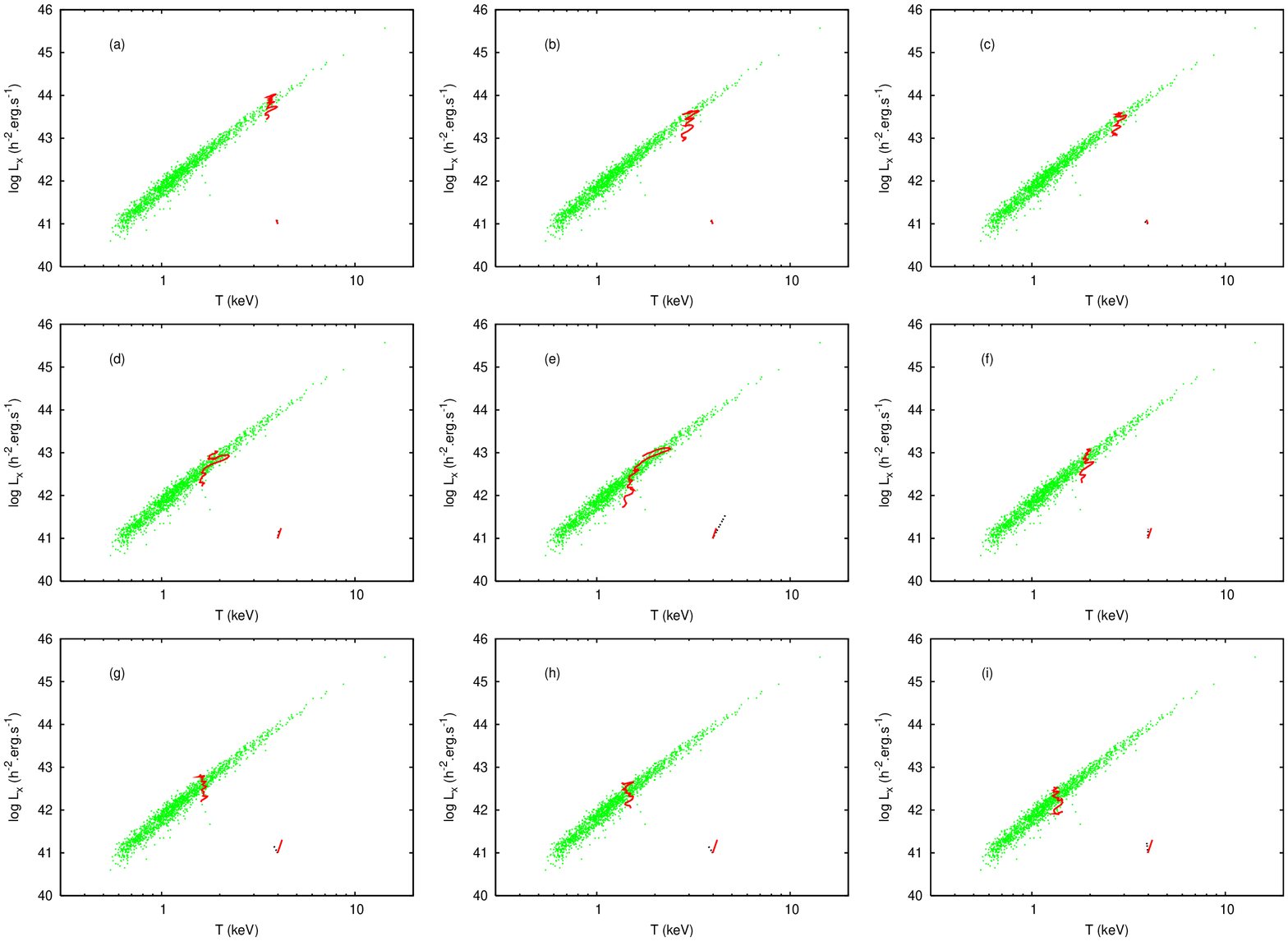}
}
\caption{L-T evolution for $9$ of the high-scattered sample. The
symbols and lines have the same meaning as in figure~\ref{llt}.}
\label{hlt}
\end{center}
\end{figure*}

\begin{figure*}
\begin{center}
\opt{bw}{
\includegraphics[angle=0, width=15cm]{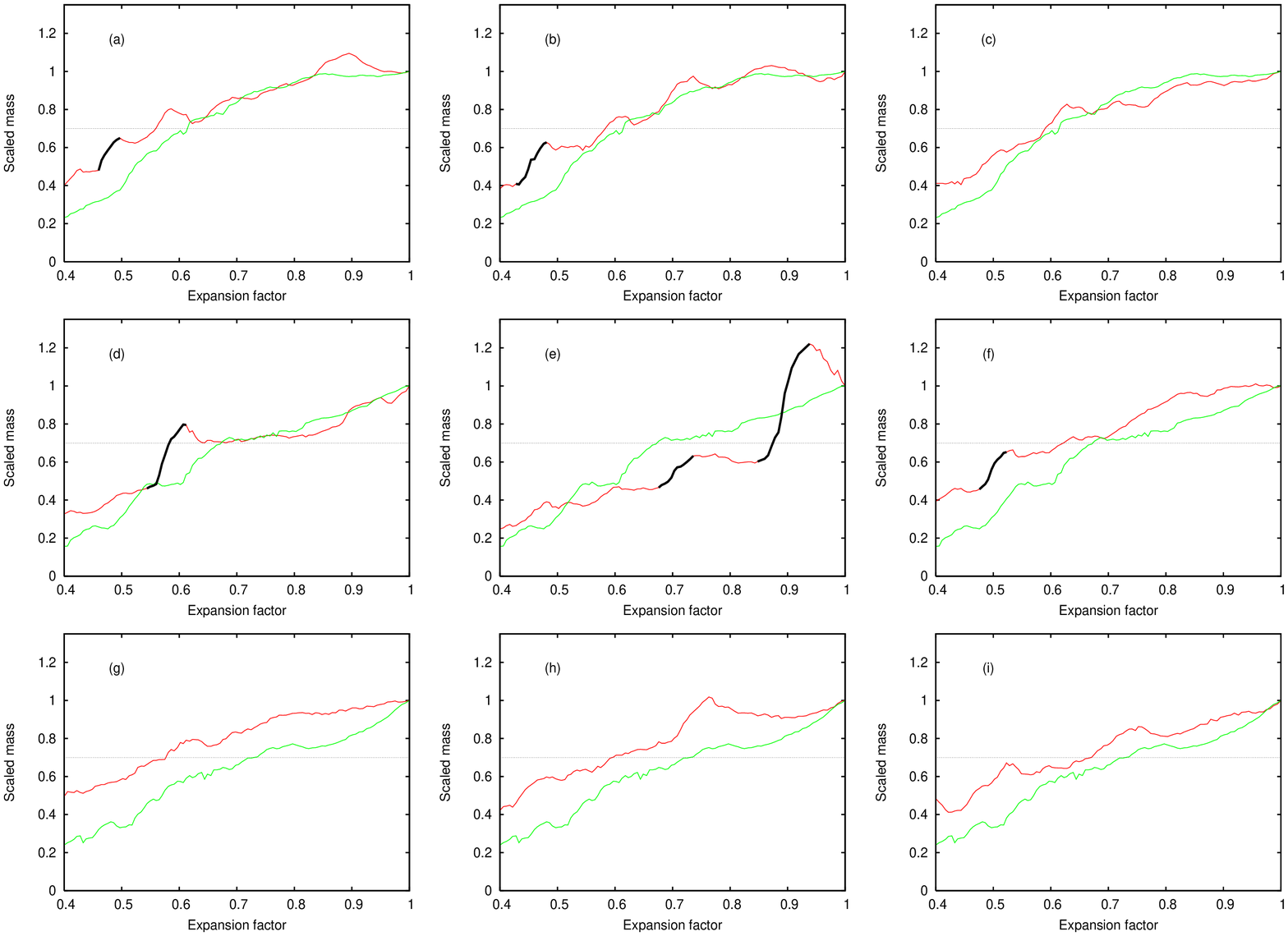}
}
\opt{col}{
\includegraphics[angle=0, width=15cm]{figs/masshigh}
}
\caption{Mass accretion histories for the high-scattered clusters in
figure~\ref{hlt}. The lines have the same meanings as in figure~\ref{lm}.
}
\label{hm}
\end{center}
\end{figure*}

Figures~\ref{hlt} \& \ref{hm} show the L-T evolution and mass
accretion histories for nine of the high scattered clusters.  These
nine halos end up significantly above the mean relation and as can be
seen from their L-T tracks in figure~\ref{hlt} eight of the nine (all
except panel e) slightly lose temperature rather than gain temperature
along with the mean of the control sample. The mass accretion history
makes it clear why this is the case: all bar panel (e) assemble their
final mass early, with significantly more mass in place at $z=1.5$
than that collected by the control sample. The object in panel (e) has
just undergone a merger.  We conclude, as did \citet{Balogh06}, that
high-scattered objects are in general early forming with consequently
slightly more concentrated dark matter profiles resulting in slightly
more luminous objects at a particular final mass.

\subsection{Properties of mergers}

As discussed in the previous section the motion of an object on the
L-T plane during a merger is a significant driver behind finding it
below the mean L-T relation at any given mass, particular at the
high-mass end.  To explore this further we extracted a sample of
mergers from the mass accretion histories of the clusters used
previously in this work. In order to distinguish a merger from gradual
accretion we require that a cluster gains significant extra mass over
a short period of time. Specifically we define a merger in the
following way:

\begin{itemize}
\item A growth in mass through the merger event of at least 15\%
of the cluster's final mass.
\item A ratio of at least $1$:$4/3$ between the mass before the merger
and the mass at the peak of the merger.
\item The mass accretion rate must exceed 14\% of the final mass per Gyr.
\end{itemize}

Merger events automatically identified using this procedure are shown
on the mass accretion history figures~\ref{lm} \& \ref{hm} as the bold
sections of the lines.  The peak of a merger is considered to be the
point at which the cluster's mass is greatest. As the dark matter
halos subsequently pass through each other the final mass is usually
below this value.  As can immediately be guessed by simply comparing the
number of bold line sections in figures~\ref{llt} \& \ref{hlt} 
the mean number of mergers undergone by clusters in
the low-scattered sample is over three times higher than clusters in
the high-scattered sample.

\begin{figure}
\begin{center}
\opt{bw}{
\includegraphics[angle=0, width=250pt]{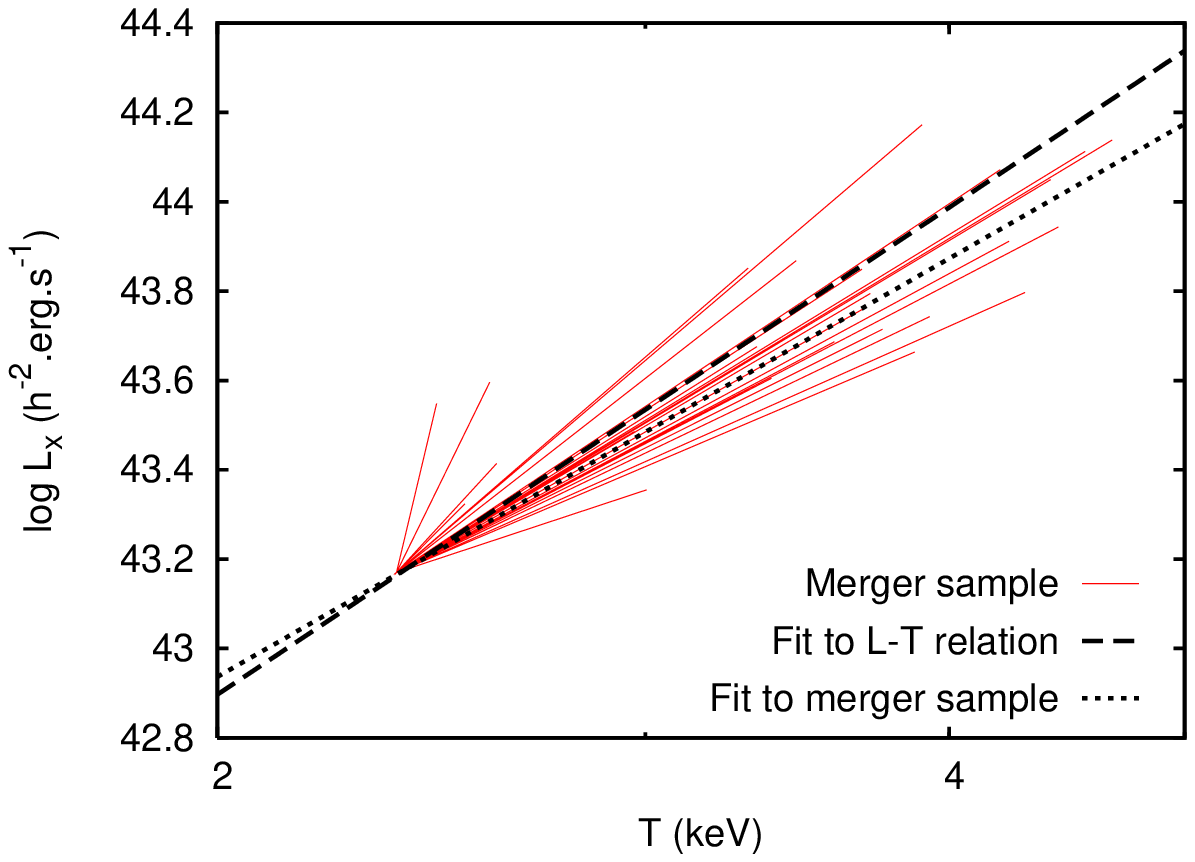}
}
\opt{col}{
\includegraphics[angle=0, width=250pt]{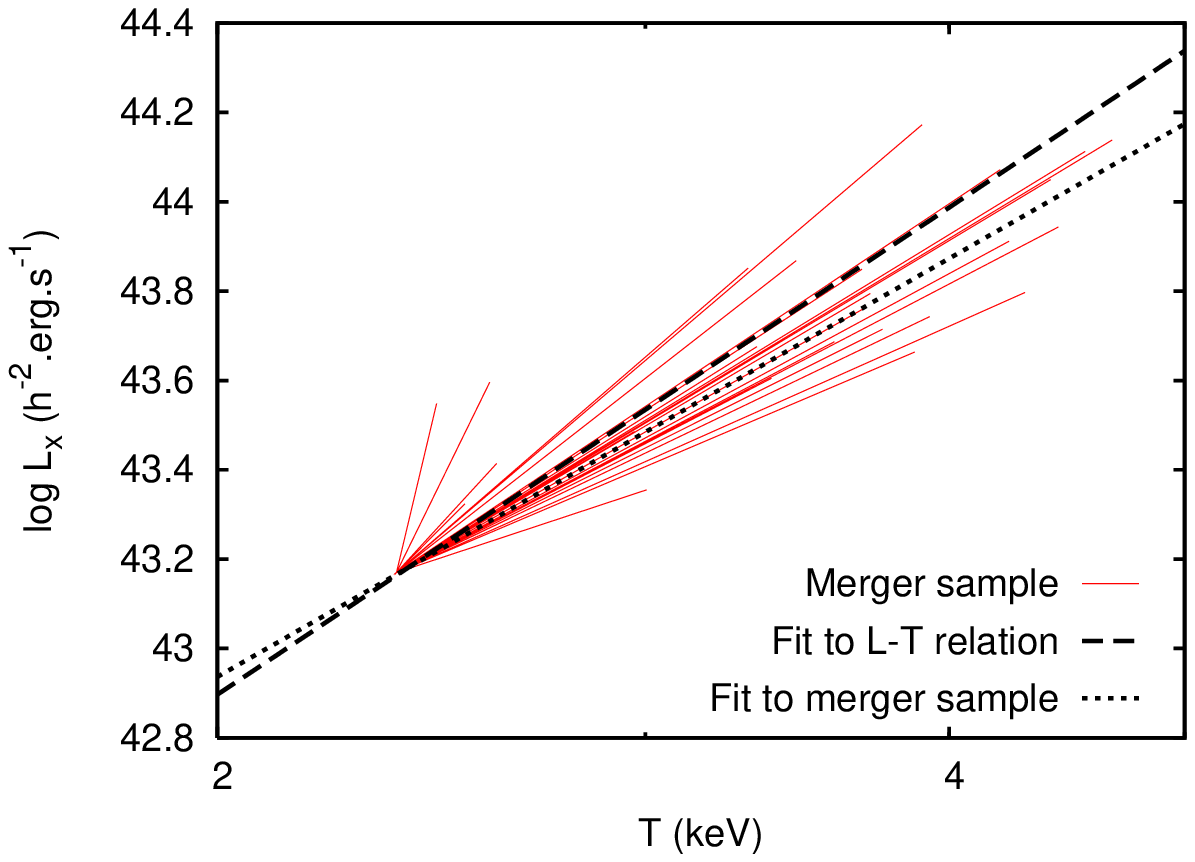}
}
\caption{Relative motion on the L--T plane during each of the mergers
  defined in section 3.3. Each line represents a single merging
  event. The long dashed line indicates the mean L--T relation whereas
  the dotted line indicates the mean merger direction. 
}
\label{mergefit}
\end{center}
\end{figure}

By identifying the location on the L-T plane of each object at the start
and peak of each merger we can produce a ``cricket score'' diagram
(figure~\ref{mergefit}), where each line denotes the motion on the L-T
plane due to one merger. As the merger timescale is short the net
drift of the relation is small while the merger is ongoing. As can be
clearly seen, the net effect of a typical merger is to move an object
up the relation and on average slightly below it. This tendency for
mergers to fall below the mean relation is further evidenced by the
large number of ongoing mergers present amongst the low-scattered
objects.

Interestingly, an ``average merger'' vector, indicated by the 
dotted line on figure~\ref{mergefit}, closely parallels the slope of
the very high mass clusters. The tendency for mergers to boost
clusters along, but at a slight angle to, the relation also drives a
slight roll that is found at the high mass end of the L-T relation.

\section{Discussion}

This work examines the physics that underlies the spine of the X-ray
L-T relation. Due to our strong preheating prescription our halos do
not have strong cores and as such do not reproduce the large scatter
in the observed L-T relation, allowing us a clear window into the
basic physics.  We intend to examine the physical origin of the
observed scatter in future work \citep{Gazzola} where a more
physically motivated energy feedback prescription will be used and
bright cooling cores are present. Preheating schemes such as the one
used here are well known to accurately reproduce the slope and
normalisation of the L-T relation as a whole \citep{Pearce, Muanwong,
Bialek,Borgani02}.  The model we have implemented also accurately
reproduces the mean location of halos on the L-T plane at the present
day but in a much larger volume than has typically been used
previously.  In the real world bright cooling cores will further
complicate matters but the processes discussed here which relate to
the outer halo properties will underlie these, with the variation in
core properties leading to a scatter about the relation discussed
here.

By identifying mergers using the mass accretion histories of our
objects and matching these episodes to the motion of each object on
the L-T plane we have derived a ``mean merger'' vector in this plane.
This vector lies largely parallel to the cluster L-T relation, as
previously noted by RO4. At any particular time the mass function of
the dark matter haloes present within a volume will be exponentially
truncated at the high mass end above some characteristic mass scale. The
large boost generated during a merger will produce points on the L-T
plane appearing to lie above this characteristic mass, where there
should be few objects. We therefore expect the majority of the
brightest objects to be experiencing ongoing mergers, although they
may be difficult to identify if they are close to their peak.

The mean merger vector we have derived is not exactly parallel to the
L-T relation but rather lies slightly below it. This behaviour leads
to all bar one of our low-scattered objects being obvious recent or ongoing
merger events (figure~\ref{lm}). We also note that at the high mass
end the vast majority of our haloes lie below the mean relation shown
on figure~\ref{sample}. The fact that the mean merger vector lies
slightly below the mean relation provides a natural explanation for
the slight curvature evidenced in the simulated relation. 

In summary, while it is straightforward to reproduce the observed
slope and normalisation of the X-ray luminosity--temperature relation
using a simple preheating scheme, such a scheme does not reproduce the
observed scatter. As a preheating model includes the full underlying
framework of the hierarchical build up of structure bulk mergers are
not significant drivers of this scatter. Mergers can, however, produce
objects that are brighter and hotter than would be expected from the
cluster mass as merger events drive objects along the L-T relation
towards the bright end. We find that a typical merger track does not
exactly parallel the L-T relation but rather lies slightly below it,
leading to a prevalence of recent or ongoing merger events on the
low-scattered side of the relation. This process also leads to a
slight curvature of the mean relation at the high mass end.

\section*{Acknowledgements}
The Millennium Simulation referred to in this paper was carried out by
the Virgo Supercomputing Consortium at the Computing Centre of the
Max-Planck Society in Garching.  The Millennium Gas Simulations were
carried out at the Nottingham HPC facility, as was the analysis
required by this work.

\label{lastpage}


\begin{thebibliography}{99}

\bibitem[\protect\citeauthoryear{Arnaud \& 
Evrard}{1999}]{Arnaud} Arnaud M., Evrard A.~E., 1999, MNRAS, 
305, 631 

\bibitem[\protect\citeauthoryear{Bialek, Evrard, \& 
Mohr}{2001}]{Bialek} Bialek J.~J., Evrard A.~E., Mohr J.~J., 
2001, ApJ, 555, 597 

\bibitem[\protect\citeauthoryear{Balogh et al.}{2006}]{Balogh06} 
Balogh M.~L., Babul A., Voit G.~M., McCarthy I.~G., Jones L.~R., Lewis 
G.~F., Ebeling H., 2006, MNRAS, 366, 624 

\bibitem[\protect\citeauthoryear{Borgani et 
al.}{2004}]{Borgani} Borgani S., et al., 2004, MNRAS, 348, 1078 

\bibitem[\protect\citeauthoryear{Borgani et 
al.}{2002}]{Borgani02} Borgani S., Governato F., Wadsley J., 
Menci N., Tozzi P., Quinn T., Stadel J., Lake G., 2002, MNRAS, 336, 409 

\bibitem[\protect\citeauthoryear{Colless et 
al.}{2001}]{Colless} Colless M., et al., 2001, MNRAS, 328, 1039 

\bibitem[\protect\citeauthoryear{Eke, Navarro, \& 
Frenk}{1998}]{ENF} Eke V.~R., Navarro J.~F., Frenk C.~S., 
1998, ApJ, 503, 569 

\bibitem[\protect\citeauthoryear{Fabian et al.}{1994}]{Fabian} 
Fabian A.~C., Crawford C.~S., Edge A.~C., Mushotzky R.~F., 1994, MNRAS, 
267, 779 

\bibitem[\protect\citeauthoryear{Faltenbacher et 
al.}{2007}]{Faltenbacher} Faltenbacher A., Hoffman Y., Gottl{\"o}ber 
S., Yepes G., 2007, MNRAS, 145 

\bibitem[\protect\citeauthoryear{Gazzola et 
al.}{2007}]{Gazzola} Gazzola, L., et al., 2007, MNRAS, submitted 

\bibitem[\protect\citeauthoryear{Helsdon \& Ponman}{2000}]{Helsdon}
Helsdon S.~F., Ponman T.~J., 2000, MNRAS, 315, 356

\bibitem[\protect\citeauthoryear{Horner et al.}{2001}]{Horner}
Horner D., Baumgartner W., Mushotzky R., Gendreau K., 2001, AAS, 1461, 33

\bibitem[\protect\citeauthoryear{Jansen et al.}{2001}]{XMM} 
Jansen F., et al., 2001, A\&A, 365, L1 

\bibitem[\protect\citeauthoryear{Kaiser}{1986}]{Kaiser} Kaiser 
N., 1986, MNRAS, 222, 323

\bibitem[\protect\citeauthoryear{Kay et al.}{2007}]{Kay07} 
Kay S.~T., da Silva A.~C., Aghanim N., Blanchard A., Liddle A.~R., Puget 
J.-L., Sadat R., Thomas P.~A., 2007, MNRAS, 234 

\bibitem[\protect\citeauthoryear{Kay, Thomas, \& 
Theuns}{2003}]{Kay03} Kay S.~T., Thomas P.~A., Theuns T., 
2003, MNRAS, 343, 608 

\bibitem[\protect\citeauthoryear{Markevitch}{1998}]{Markevitch} 
Markevitch M., 1998, ApJ, 504, 27 

\bibitem[\protect\citeauthoryear{McCarthy et 
al.}{2004}]{McCarthy} McCarthy I.~G., Balogh M.~L., Babul A., 
Poole G.~B., Horner D.~J., 2004, ApJ, 613, 811 

\bibitem[\protect\citeauthoryear{Muanwong et 
al.}{2001}]{Muanwong} Muanwong O., Thomas P.~A., Kay S.~T., 
Pearce F.~R., Couchman H.~M.~P., 2001, ApJ, 552, L27 

\bibitem[\protect\citeauthoryear{Mulchaey et al.}{2003}]{Mulchaey}
Mulchaey J.~S., Davis D.~S., Mushotzky R.~F., Burstein D., 2003, ApJS, 145, 39M

\bibitem[\protect\citeauthoryear{Navarro, Frenk, \& 
White}{1995}]{NFW} Navarro J.~F., Frenk C.~S., White 
S.~D.~M., 1995, MNRAS, 275, 720 

\bibitem[\protect\citeauthoryear{Navarro, Frenk, \& 
White}{1997}]{NFW97} Navarro J.~F., Frenk C.~S., White 
S.~D.~M., 1997, ApJ, 490, 493 

\bibitem[\protect\citeauthoryear{Osmond \& 
Ponman}{2004}]{Osmond} Osmond J.~P.~F., Ponman T.~J., 2004, 
MNRAS, 350, 1511 

\bibitem[\protect\citeauthoryear{Pearce et 
al.}{2007}]{Pearce07} Pearce, F. R., et al., 2007, MNRAS, submitted 

\bibitem[\protect\citeauthoryear{Pearce et al.}{2000}]{Pearce} 
Pearce F.~R., Thomas P.~A., Couchman H.~M.~P., Edge A.~C., 2000, MNRAS, 
317, 1029 

\bibitem[\protect\citeauthoryear{Pierre et al.}{2006}]{Pierre}
Pierre M., et al., 2006, MNRAS, 372, 591 

\bibitem[\protect\citeauthoryear{Poole et al.}{2006}]{Poole}
Poole G.~B., et al., 2007, MNRAS, 380, 437P

\bibitem[\protect\citeauthoryear{Ritchie \& Thomas}{2002}]{Ritchie}
Ritchie B.~W., Thomas P.~A., 2002, MNRAS, 329, 675R

\bibitem[\protect\citeauthoryear{Romer et al.}{2001}]{Romer}
 Romer A.~K., Viana P.~T.~P., Liddle A.~R., 
2001, ApJ, 547, 594R

\bibitem[\protect\citeauthoryear{Rowley, Thomas, \& 
Kay}{2004}]{Rowley} Rowley D.~R., Thomas P.~A., Kay S.~T., 
2004, MNRAS, 352, 508 [R04]

\bibitem[\protect\citeauthoryear{Sarazin}{1986}]{Sarazin} 
Sarazin C.~L., 1986, RvMP, 58, 1 

\bibitem[\protect\citeauthoryear{Schwope et al.}{2003}]{Schwope} 
Schwope A. D., Lamer G., Burke D., Elvis M., Watson M.~G.,
Schulze M.~P., Szokoly G., Urrutia T., 2004, AdSpR, 34, 2604

\bibitem[\protect\citeauthoryear{Spergel et 
al.}{2003}]{Spergel} Spergel D.~N., et al., 2003, ApJS, 148, 
175 

\bibitem[\protect\citeauthoryear{Springel et 
al.}{2005}]{Springel} Springel V., et al., 2005, Natur, 435, 629 

\bibitem[\protect\citeauthoryear{Sutherland \& 
Dopita}{1993}]{Sutherland} Sutherland R.~S., Dopita M.~A., 1993, 
ApJS, 88, 253 

\bibitem[\protect\citeauthoryear{Weisskopf et 
al.}{2002}]{Chandra} Weisskopf M.~C., Brinkman B., Canizares 
C., Garmire G., Murray S., Van Speybroeck L.~P., 2002, PASP, 114, 1 

\bibitem[\protect\citeauthoryear{Wu et al.}{1999}]{Wu99}
Wu X., Xue Y., Fang L., 1999, ApJ, 524, 22

\bibitem[\protect\citeauthoryear{Xue \& Wu}{2000}]{Xue}
Xue Y., Wu, X., 2000, ApJ, 538, 65

\end{thebibliography}
\end{document}